\documentstyle[12pt,epsf,rotate]{article}
\makeatletter

\@addtoreset{equation}{section}
\makeatother
\begin{document}
\title{Variation of transport coefficients for average
fission dynamics with temperature and shape} 
\author{Shuhei Yamaji$^{1}$,  Fedor A.Ivanyuk$^{2,3}$ and Helmut Hofmann $^{2}$
\\
\small\it{ 1) Cyclotron Lab., Riken, Wako, Saitama, 351-01, Japan
\thanks{yamajis@rikaxp.riken.go.jp}
}
\\
\small\it{2) Physik-Department der Technischen Universit\"at M\"unchen,
D-85747 Garching, Germany
\thanks{hhofmann@physik.tu-muenchen.de \hfill
http://www.physik.tu-muenchen.de/tumphy/e/T36/hofmann.html}
}
\\
\small\it{3) Institute for Nuclear Research of the Ukrainian Academy of
Sciences, Kiev-28, Ukraine}}
\date{}
\maketitle
\begin{abstract}

We study slow collective motion at finite thermal excitations on the basis
of linear response theory applied to the locally harmonic approximation.
The transport coefficients for average motion, friction $\gamma$,
inertia $M$ and the local stiffness $C$ are computed along a fission
path of $^{224}Th$ within a quasi-static picture. The inverse relaxation
time $\beta=\gamma/M$ and the effective damping rate
$\eta=\gamma/(2\sqrt{M|C|})$ are found to increase with temperature,
but do not change much with the collective variable. The values found
for $\eta$ and $\beta$ as well as their behavior with temperature are in
accord with experimental findings.
\end{abstract}

\vskip 0.5cm
\centerline{PACS: -05.60.+w, 21.60.Cs, 21.60.Ev, 24.10Pa, 24.75+i}
\vskip 0.5cm
\centerline{\bf Nucl. Phys. A612 (1997) 1}
\vskip 0.5cm

\section{Introduction}

The nature of nuclear dissipation is not well understood as yet. 
For isoscalar modes at finite excitations the best information
available presently comes from fission experiments, when comparing the
decay rate of fission with the ones for emission of light particles or
gamma rays. Nowadays it seems not only possible to deduce numbers for the
effective damping rate $\eta$  \cite{hirorev}, \cite{higoroobn} but to
gain information about its temperature dependence as well \cite{pauthoe}.
The authors of \cite{pauthoe} find an $\eta$ which increases markedly
with $T$, at small to moderately large values of the temperature. 

Such behavior is hard to understand for macroscopic models
for which friction is either treated on the basis of two body
viscosity or on the wall formula \cite{bbnrrss}, \cite{swiatcop}. In
the first case friction should decrease with $T$ as $T^{-2}$, in the
second it would practically stay constant. Both results represent
two contrasting pictures of the nuclear dynamics. The first one assumes
collision dominance, and is thus related in a sense to the nuclear
compound model or its classical analog, the liquid drop model. The
second picture would apply if, between two encounters or "collisions"
with the wall, the nucleons are allowed to move  freely inside the
nuclear medium, as is suggested by the models of independent
particles. In principle, the other two transport coefficients appearing
in $\eta$, namely $M$, $C$ must be expected to change with temperature,
too. But for the macroscopic models just mentioned this dependence will
be weak. Therefore, the observed increase of $\eta$ with $T$ gives
strong evidence for the necessity of having a microscopic theory. One
possible formulation of the problem and its solution is based on a
specific application of linear response theory. For a detailed
description of the latter we may refer to 
\cite{yahosa} -
\cite{hoivya} and \cite{ho}.
In the present paper this theory will be applied to continue previous 
microscopic computations of the coefficients for friction $\gamma$,
inertia $M$ and local stiffness $C$, such as the ones of \cite{yahosa}
and \cite{hoivya}. Emphasis will be put on their dependence both on
temperature $T$ as well as on the collective coordinate $Q$. 

\section{The linear response approach to collective motion}

In this section we only want to outline briefly some basic
theoretical features to have the most relevant formulas ready for
explaining our computational analysis; we will follow 
largely \cite{kidhofiva}, \cite{hoivya} and \cite{ho}. 

Similarly to the deformed shell model, we assume to have at our
disposal a Hamiltonian $\hat{H}(\hat{x}_i, \hat{p}_i,Q)$ which depends
on deformation through the shape variable $Q$. For the sake of
simplicity we just take one such degree of freedom, which later-on
will be used to parameterize motion along some given fission path.
However, it does not suffice to restrict this $\hat{H}(\hat{x}_i,
\hat{p}_i,Q)$ to the model of independent particles, for which one
would have (for A nucleons)
\begin{equation}
\hat H_{sm}(\hat x_i, \hat p_i, Q) =\sum_{l=1}^A 
\hat h(\hat {\vec x}_l, \hat {\vec p}_l, Q)\quad, 
\label{hmf}
\end{equation}
where $\hat h(\hat {\vec x}_l, \hat {\vec p}_l, Q)$ stands for the
dynamics of particle l. First of all, the
expectation value of $H_{sm}(\hat x_i, \hat p_i, Q)$ does not represent
the system's total  energy. Secondly, as it stands this Hamiltonian
does not account for the effects of collisions. Following \cite{sije} the
first deficiency is easily cured by adding to the operator part 
$\hat H_{sm}(Q)$ a c-number term to get a $\hat H_{rmf}(Q)$ as the
representative for the (renormalized) mean field. This c-number term
can be chosen in such a way that the static expectation value of $\hat
H_{rmf}(Q)$ (i.e. the one calculated for a time independent $Q$)
contains the re-normalization terms of the Strutinsky method. (We choose
the notation of \cite{ho} which differs slightly from \cite{sije}, but
the connection between both is readily established). The second issue
can be taken care of by adding the effects of collisions when treating
dynamical forces. In principle one might write
\begin{equation}
\hat H(\hat x_i,\hat p_i, Q)= 
\hat H_{rmf}(Q) +\hat V^{(2)}_{res}(\hat x_i,\hat p_i)    
\label{hamilt}
\end{equation}
with the restriction of having the residual two body interaction 
$\hat V^{(2)}_{res}(\hat x_i,\hat p_i)$ appear
only in dynamical quantities like the response functions. How
that can be done in practice will be explained below. Here we would
just like to mention that we want to assume this {\it incoherent}
interaction $\hat V^{(2)}_{res}(\hat x_i,\hat p_i)$ to be {\it
independent} of the collective coordinate $Q$.

With these precautions taken into account we may say that
the average $\langle\hat{H}(\hat{x}_i,\hat{p}_i,Q)\rangle$ represents
the total energy $E_{tot}(t)$ of the system, and now even in a dynamical
sense. However, in case that the nucleus is {\it isolated} this energy
must be a {\it constant of motion}. Hence the equation of motion for the
$Q(t)$ can be constructed from energy conservation by applying Ehrenfest's
theorem \cite{holet}. This is to say we may write:
\begin{equation}
0 = \frac{d}{dt}E_{tot} = \dot{Q}\langle\frac{\partial
\hat{H}(\hat{x}_i,\hat{p}_i,Q)}{\partial Q}\rangle _t \equiv \dot{Q} \langle
\hat{F}(\hat{x}_i,\hat{p}_i,Q)\rangle _t
\label{encon}
\end{equation}
The remaining task then is to
express the average $\langle\hat{F}(\hat{x}_i,\hat{p}_i,Q)\rangle_t$
as a functional of $Q(t)$. It is here that we shall exploit the
benefits of linear response theory. The relevant operator
$\hat{F}(\hat{x}_i,\hat{p}_i,Q)$ is seen to be
given by the derivative of the mean field with respect to $Q$; it is of
a purely one body nature as the residual interaction $\hat V^{(2)}_{res}(\hat
x_i,\hat p_i)$ drops out when calculating this derivative. 

\subsection{Local linearization}

Under certain circumstances, the task of evaluating the functional form of 
$\langle\hat{F}(\hat{x}_i,\hat{p}_i,Q)\rangle_t$ can be considerably
simplified. This will be so whenever the relevant $Q$  can be handled
as being close to some fixed value $Q_0$. Quite naturally this would be
the case for harmonic vibrations about a stable potential minimum at
$Q_0$.  Fortunately, such a situation may be given even in the more
general case. In this paper we restrict ourselves to study average
dynamics, which means to neglect any statistical fluctuations in the
variable $Q$, in which case it suffices to require collective motion to
be sufficiently slow. Then $Q$ will stay in the neighborhood of some
given $Q_0$ for a ("microscopically") large time interval $\delta t$;
if necessary one may interpret the $Q_0$ as the $Q(t)$ at a given time
$t_0$.

For any $Q$ close to $Q_0$ one may evaluate the {\it intrinsic} quantity 
$\langle\hat{F}(\hat{x}_i,\hat{p}_i,Q)\rangle_t$ by effectively using 
the  Hamiltonian
\begin{equation}
\hat{H}(\hat{x}_i,\hat{p}_i,Q) = \hat{H}(\hat{x}_i,\hat{p}_i,Q_0) + 
(Q - Q_{0})\hat{F}(\hat{x}_i,\hat{p}_i,Q_0) + \frac{1}{2}(Q - Q_{0})^2
\langle\frac{\partial^{2}\hat{H}}{\partial Q^{2}} (Q_{0})\rangle
^{qs}_{Q_{0},T_{0}} \quad ,
\label{hamiltonian}
\end{equation}
instead of the original one. It is obtained by expanding the
Hamiltonian to second order and by approximating the second order term
in a kind of unperturbed limit. By this we mean to evaluate the
expectation value on the very right by a {\it static} density operator
$\hat{\rho}_{qs}=\hat{\rho}_{qs}\{\hat{H}(Q_0)\}$ which is determined 
by the Hamiltonian $\hat{H}(\hat{x}_i,\hat{p}_i,Q_0)$ taken at $Q_0$.
Effectively, the only coupling term left between collective and
intrinsic motion is then given by the term of first order in $Q-Q_0$.
It is not difficult to grasp the concept behind such an approximation:
In this way global motion is described within a locally harmonic
approximation. Here it was developed for average motion; a discussion
of the general case can be found in \cite{ho}.  One last remark on
notation: In the sequel the $\hat{F}(\hat{x}_i,\hat{p}_i,Q)$ will be
denoted by $\hat{F}$ whenever it is to be taken at $Q_0$.

Applying this concept to derive the equation of motion, it can be shown
(see e.g. \cite{kidhofiva} and \cite{ho}) that eq.(\ref{encon}) leads to the
following form:
\begin{equation}
k^{-1}q(t) + \int_{-\infty }^{\infty}\tilde{\chi}(s)q(t-s)ds = 0.
\label{eom}
\end{equation}
Here $q=Q-Q_{m}$ measures the deviation of the actual $Q$ from the
center of the oscillator approximating the true potential in the
neighborhood of $Q_{0}$. The $\tilde{\chi}$ is the causal response
function associated to the dynamics of the nuclear ``property'' 
$\langle\hat{F}\rangle $. It is given by
\begin{eqnarray}
&\tilde{\chi}(t-s)=\Theta(t-s)\frac{i}{\hbar}
tr(\hat{\rho}_{qs}(Q_{0},T_{0})[\hat{F}^{I}(t),\hat{F}^{I}(s)])\nonumber
\\
&\equiv 2i\Theta(t-s)\tilde{\chi}^{\prime\prime}(t-s)
\label{respt}
\end{eqnarray}
The expectation value appearing here is to be calculated like the one
encountered in (\ref{hamiltonian}), namely by the density
operator $\hat{\rho}_{qs}\{\hat{H}(Q_0)\}$. The same Hamiltonian
appearing there is used to specify the time evolution in $\hat{F}^I(t)$. The
$\hat{\rho}_{qs}$ is meant to represent a thermal equilibrium at $Q_0$
with excitation being parameterized either by temperature or by entropy. The
quantity $k$ summarizes contributions of static forces which appear in
second order. Anticipating the change in entropy to be quadratic in
$\dot{q}(t)$, one gets for the coupling constant $k$:
\begin{equation}
- k^{-1} = \langle \frac{\partial ^2 \hat{H}}{\partial Q^2} (Q_0)
\rangle^{qs}_{Q_{0},T_{0}} + (\chi (0) - \chi ^{ad})
\label{coupling}
\end{equation}
with $\chi(0)$ being the static response (the Fourier transform of the
time-dependent response function (\ref{respt}) taken at frequency $\omega
=0$) and $\chi^{ad}$ being the adiabatic susceptibility (sometimes
$\chi(0)$ is referred to as isolated susceptibility). 
It was shown in \cite{kidhofiva} that for temperatures not smaller than
1 to $1.5 MeV$ the coupling constant may effectively be calculated
from the free energy $f$, or its stiffness ${\partial^{2}f}/{\partial
Q^{2}_{0}}$, rather, using the formula 
\begin{equation}
-k^{-1}\simeq -k^{-1}\Bigr\arrowvert_{T=const}=\frac{\partial^{2}f}{\partial
Q^{2}_{0}}\Bigr\arrowvert_{T}+\chi(0)
\label{coupling3}
\end{equation}

\subsection{Transport coefficients from the collective response}

Transport coefficients parameterize properties of a system whose time
development is described by differential equations. The easiest way of
getting such an equation from the integral form (\ref{eom}) is to expand
the factor $q(t-s)$ under the integral to second order in
$s$ \cite{holet}. In this way the common equation of motion of the damped 
oscillator
is obtained 
\begin{equation}
M \ddot q (t) + \gamma \dot q(t) + C q(t) = 0
\label{oscequat}
\end{equation}
and the transport coefficients attain the forms:
\begin{equation}
M = M(0)=-{1\over 2}\int_{-\infty}^{\infty} \tilde \chi(s)s^2~ds
=\frac{1}{2}\frac{\partial^{2}\chi}{\partial\omega^{2}}
\Bigr\arrowvert_{\omega=0}
\label{mass0}
\end{equation}
\begin{equation}
\gamma =\gamma(0)=\int_{-\infty}^{\infty} \tilde \chi(s)s~ds=
\frac{\partial\chi^{\prime\prime}}{\partial\omega}\Bigr\arrowvert_{\omega=0}.
\label{gamma0}
\end{equation}
\begin{equation}
C=C(0)=-\chi (0)-k^{-1}=\frac{\partial^{2}f}{\partial Q^{2}_{0}}
\Bigr\arrowvert_{T}
\label{c0}
\end{equation}
They follow after evaluating the moments in time
of the response function by Fourier transforms. 
The last equality in eq.(\ref{c0}) is a consequence of the expression 
(\ref{coupling3}) for the coupling constant. 
For obvious reasons the notion "zero-frequency limit" has been coined
to portray the coefficients (\ref{mass0}- \ref{c0}). 

We may note that the procedure just applied is borrowed from models
where the basic equations of motion are of integro-differential type,
and where the reduction to differential form commonly is referred to as
Markov approximation. It is only because of the self-consistency
underlying the derivation sketched above that in (\ref{eom}) the common
inertial terms are missing. As typically they are of order zero in the
coupling they would make up differential terms from the start.  Unfortunately,
for the present case the method behind (\ref{oscequat}- \ref{c0})
does not always lead to a decent approximation to (\ref{eom}). Let us
mention just two problems. The most stringent one can be seen in the
fact that the expression for the inertia, which is nothing else but an
extension of the cranking inertia to the case of damped motion at
finite temperature \cite{holet}, may become negative. Secondly, as both
(\ref{eom}) as well as (\ref{oscequat}) are homogeneous equations, the
transport coefficients obtained in this way are defined up to a common
factor only. To obtain all three coefficients, namely inertia $M$,
friction $\gamma$ and stiffness $C$, one needs additional information.

Following general concepts, this information can be obtained from a
response function which parameterizes {\it collective} motion locally,
and which will thus be denoted by $\chi_{coll}(\omega)$. It can
be derived by introducing a (hypothetical) external force
$\tilde{f}_{ext}(t)\hat{F}$ and by evaluating how the deviation  of
$<\hat{F}>_{\omega}$ from some properly chosen static value reacts to
this external field in linear order: $\delta
<\hat{F}>_{\omega}=-\chi_{coll}(\omega)f_{ext}(\omega)$. As shown in
\cite{kidhofiva} and \cite{ho} the $\chi_{coll}(\omega)$ can be brought
to the form
\begin{equation}
\chi_{coll}(\omega)=\frac{\chi(\omega)}{1+k\chi(\omega)}
\label{collresp}
\end{equation}
which is known to be standard for the case of zero temperature (see
e.g. \cite{bomo} and \cite{sije}). For finite excitations one needs an
additional condition, namely that the motion is ergodic,
in the sense \cite{brenigtwo} of having adiabatic and isolated
susceptibility equal to each other: $\chi (0) = \chi ^{ad}$.

The dissipative part of $\chi_{coll}(\omega)$ represents the
distribution of strength over various possible local modes, which
exhibit themselves as individual peaks. The corresponding "dispersion
relation" or secular equation 
\begin{equation}
\frac{1}{k} + \chi(\omega) =0 
\label{dispers}
\end{equation}
is easily recognized as the Fourier transform of (\ref{eom}).  From
(\ref{dispers}) one gets a series of complex frequencies $\omega_\nu $,
which actually come in pairs $\omega_\nu^\pm = \pm {\cal E}_\nu - i
\Gamma_\nu /2$. Each pair may be associated to the solutions of the
secular equation of a damped oscillator like (\ref{oscequat}). Any
$q(t)$ satisfying (\ref{eom}) fully or correctly will contain information
about {\it all} the  $\omega_\nu $. However, restricting oneself to
just one such pair the integral equation effectively are reduced to a
differential form.  Since we aim at describing {\it slow} collective
motion, it is natural to take the lowest ones, say $\omega_1^\pm$. In
addition to the two parameters specifying them, we have as additional
information the strength of the corresponding poles. This feature
enables one to evaluate all three transport coefficients $M, \gamma$
and $C$. In practice this can be done by replacing
\begin{eqnarray}
\label{oscresp}
(\chi_{coll}(\omega))^{-1}\delta<F>_{\omega} & = &
-f_{ext}(\omega) \nonumber \\
&\Downarrow &  \\ 
(\chi_{osc}(\omega))^{-1}\delta<F>_{\omega}
&\equiv & 
k^{2}(-M\omega^{2}-\gamma i \omega+C)\delta<F>_{\omega}
=-f_{ext}(\omega), \nonumber 
\end{eqnarray}
which is to say by approximating the response associated of the low
frequency mode by that of a damped oscillator with modified transport
coefficients. In practice the latter can be found
by fitting the dissipative part of the oscillator response to the peak
in the original strength distribution
$\chi_{coll}^{\prime\prime}(\omega)$.

The appearance of the factor $k^2$ has a simple mathematical reason.
Evaluating the contribution to the collective response from the two
poles chosen, the numerator of (\ref{collresp}) has to be calculated at
the poles' frequencies, which in the $\chi_{coll}(\omega)$
leads to the factor $\chi(\omega_\nu^\pm)=-1/k$. Please notice that
by construction the oscillator response introduced here is the one for
the quantity $<F>_t$, while the transport coefficients written in
(\ref{oscresp} are those for the "Q-mode". Indeed, the 
form (\ref{collresp}) requires that (on average) motion in $F$ is related to
that in $Q$ by $k<F>_t = Q(t) - Q_0$. Therefore, the corresponding
transport coefficients must be related to each other like  ${\cal T}^F
= k^2 {\cal T}$, where ${\cal T}$ stands for $M,\gamma,C$.

As just indicated, the coefficients introduced in this way reflect the
structure of the collective response (\ref{collresp}), for instance by
way of the self-consistency between collective and intrinsic motion
\footnote{Please recall that for undamped motion (\ref{collresp}) is
identical to the response function of RPA one gets for the separable
interaction $(k/2) \hat F \hat F$.}
. Therefore, we will at times call these coefficients the
self-consistent ones. 
It is possible to relate them to the zero-frequency limit. Suppose
the lowest pair of frequencies $\omega_1^\pm$ lies sufficiently close
to $\omega =0$. Then the $\omega_1^\pm$  may be obtained by expanding
in (\ref{dispers}) the $\chi(\omega)$ to second order in $\omega$
around $\omega=0$. The result is easily recognized as the secular
equation to (\ref{oscequat}) with the transport coefficients being
given by (\ref{mass0} - \ref{c0}). In \cite{hosaya} some conditions
have been derived for the zero-frequency limit to apply,
for the case where the  collective response function
$\chi_{coll}(\omega)$ is well simulated by the oscillator response
function $\chi_{osc}(\omega)$, i.e. where the reduction (\ref{oscresp})
does not imply any further approximation. From the equations
shown in \cite{hosaya}  it is easy to see that the following relations
hold true:
\begin{equation}
C=C(0)\left(1+\frac{C(0)}{\chi(0)}\right)
\label{capprox}
\end{equation}
\begin{equation}
\gamma=\gamma(0)\left(1+\frac{C(0)}{\chi(0)}\right)^{2}
\label{gmapprox}
\end{equation}
\begin{equation}
 M= \left({C\over C(0)}\right)^2\left(M(0)+{\gamma^2(0)\over \chi (0)}\right). 
\label{msapprox}
\end{equation}
Once more, the transport coefficients of the left hand side are the ones 
obtained for the fit and those of the right hand side represent the
zero frequency limit of the Lorentzian defined by the $M,\gamma, C$.
Please observe, that the form (\ref{collresp}) may be turned around to
express $\chi(\omega)$ by  $\chi_{coll}(\omega)$. Replacing the latter
by $\chi_{osc}(\omega)$ the functional form of $\chi(\omega)$ is 
seen to be of Lorentzian type as well.

\begin{figure}[hb]
\centerline{{\rotate[r]{\epsfysize=12cm \epsffile{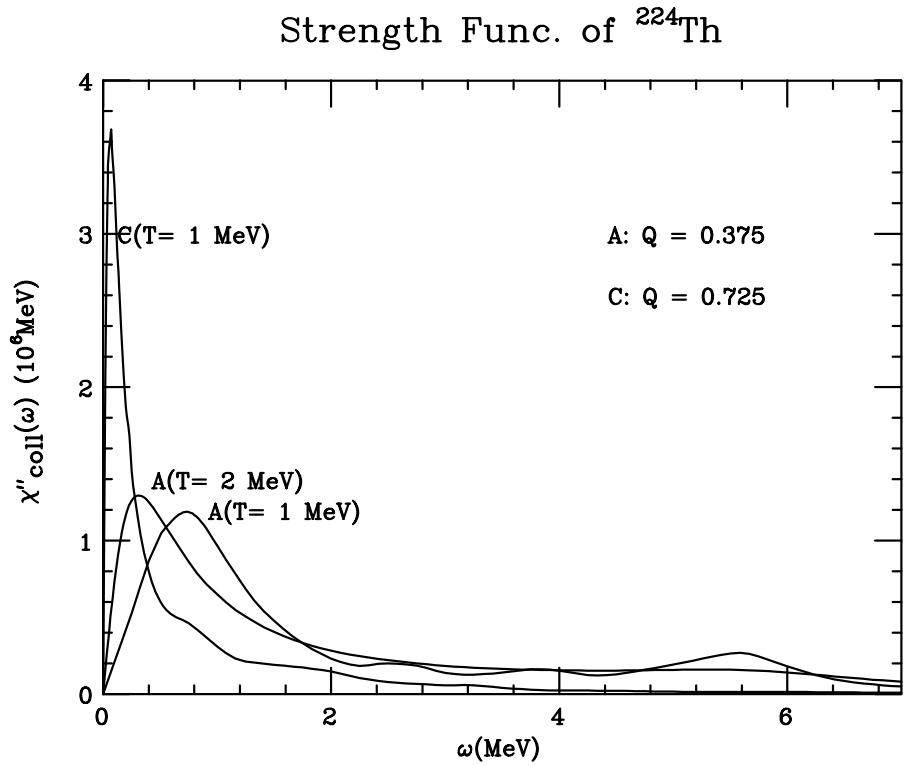}}}}
\caption{The imaginary part of the collective response function (in
arbitrary units) as function of $\omega$, calculated at the points A
and C for T = 1 and 2 MeV.}
\label{imagcoll}
\end{figure}

Before we continue we would like to demonstrate that this concentration
in a low frequency peak can be considered realistic for many situations,
provided the temperature is not too small. In Fig.\ref{imagcoll} we
show for temperatures $T = 1$ and $2 \;MeV$ the strength function
$\chi^{\prime\prime}_{coll}(\omega)$ at the potential minimum and at
the saddle point, which later on will be called point A and C,
respectively. (Details of the computation will be explained in the next
section). Whereas at A the strength distribution reflects the typical
behavior of stable modes, the one at C corresponds to unstable motion.
For stable motion peaks are seen at higher frequencies, but the latter
get washed out with increasing excitation. These features have been
recognized and discussed before in \cite{hoyaje} for quadrupole
vibrations of $^{208}Pb$. The resulting concentration in a low
frequency mode is even more clearly seen at the instability, where it
already occurs at the smaller temperature of  $T = 1 MeV$.  We should
like to stress that this somewhat peculiar behavior of having the main
peak appear at very low frequencies is strongly related to the fact that
the motion is over-damped. This can be made quantitative for
the case of the oscillator response introduced in (\ref{oscresp}). It
is easy to convince one-selves that the position $\omega_m$ of the
maximum of $\chi^{\prime\prime}_{osc}(\omega)$ is found at
\begin{equation}
\omega_m^2=\frac{\varpi^2}{3}
\left(2\sqrt{1-{\rm sign}C\eta^2+\eta^4}+{\rm sign}C-2\eta^2\right)
\label{ommax}
\end{equation}
where $\varpi^2=\vert C \vert /M$. In the limit of $\eta \to \infty$
$\omega_m$ turns to $|C|/\gamma$ and the value of the response function
taken at $\omega_m $ is identical to $\chi_{osc}(\omega_m)=1/(2k^{2}|C|)$.

Finally we like to put Eq-s.(\ref{capprox} to \ref{msapprox}) on more
general grounds. Before they have been derived on the basis of the
Lorentzian fit to the collective response function (\ref{collresp}). 
However, quite generally, one may start with the coefficients of the
zero frequency limit obtained from the full, original intrinsic response
function. These coefficients come into play expanding again to second
order in $\omega$ around  $\omega=0$, this time, however, not the 
$\chi (\omega)$ itself, but the $(\chi_{coll}(\omega))^{-1}= k + (\chi
(\omega))^{-1}$ instead. In this way approximate versions of the
self-consistent transport coefficients are obtained. It so turns out
that they obey the same relation to the zero-frequency limit as given
by (\ref{capprox} - \ref{msapprox}).  Later on we shall compare
numerically the $M, \gamma, C$ of (\ref{capprox} - \ref{msapprox}) with
the corresponding coefficients obtained by the fit indicated in
(\ref{oscresp}). We may indicate already here that typically
${C(0)}/{\chi(0)}$ is very small and decreases with temperature.  This
can be inferred from the fact that $C(0)$ drops with increasing
excitation whereas $\chi(0)$ turns out quite insensitive to changes in
$T$ (c.f.\cite{kidhofiva}). For realistic cases, say above $T\approx 1
MeV$, ${C(0)}/{\chi(0)}$ is of the order of several \%. For this reason
the $C$ and the $\gamma$ of (\ref{capprox}) and (\ref{gmapprox}) get
close to their zero-frequency value. It is only the inertia $M$ of
(\ref{msapprox}) which may differ considerably from $M(0)$, as a
consequence of the second term in the second bracket. It is interesting
to note, that at $T=0$ the ratio ${C(0)}/{\chi(0)}$ might become of
order unity if the spin-orbit interaction would not play a crucial
role. This fact may be inferred from the analysis of quadrupole
vibrations presented in sect.8.5 of
\cite{sije} (from eq.(8.5.13b) it can be deduced that in this case
${C(0)}/{\chi(0)}$ becomes identical to unity).

\subsection{Collisional damping of nucleonic motion}

Let us turn to the damping mechanism used in our theory. From the
discussion of (\ref{hamilt}) one may anticipate that finally it is 
the residual two-body interaction
$V^{(2)}_{res}(\hat{x}_{i},\hat{p}_{i})$ which causes damping, first on
the microscopic level, and then by way of self-consistency for the
collective motion as well. Because of the assumption of this
interaction being independent of $Q$ it enters the game only through
the Hamiltonian $\hat{H}(\hat{x}_i,\hat{p}_i,Q_0)$ which explicitly
appears in the response function (\ref{respt}).  To evaluate this
expressions fully for some given $V^{(2)}_{res}$ would be too a
tremendous task. We therefore use a scheme which borrows from the way
one would treat the effects of collisions in time dependent mean field
theories like ETDHF or its classic versions as given by the BUU or
Landau-Vlasov equation. In this paper we will just state the final
expressions referring both to \cite{ho} as well as to earlier
publications (for detailed lists see \cite{yahosa} and \cite{hoivya}).

The Fourier transform of the dissipative part of the
intrinsic response function finally writes like
\begin{equation}
\label{response}
   \chi^{\prime \prime }(\omega)  = 
   \int\;{d\hbar\Omega\over 4\pi}\;
   \Bigl(n(\Omega-{\omega\over 2}) -n(\Omega+
    {\omega\over 2}) \Bigr) 
   \sum _{jk } \big\arrowvert F_{jk }\big\arrowvert ^2  
   \varrho_k(\Omega - {\omega\over 2})  
    \varrho_j(\Omega  + {\omega\over 2})
\end{equation}
Here, $n(x)$ is the Fermi function determining the occupation of the
single particle levels $|k>$. The latter are the eigenstates of the 
Hamiltonian $\hat{h}(\hat {\vec x}, \hat {\vec p}, Q_{0})$ with
corresponding energies $e_{k}$. 
The $\varrho_k(\omega) $ represents the
distribution of the single particle strength over more complicated
states. It is here that the effects of collisions come into play.
Neglecting them, which in our present language means to say 
putting $V^{(2)}_{res}$ equal to zero, the $\varrho_k(\omega) $ would
simply be given by $\varrho_k(\omega) = 2\pi \delta(\hbar \omega -
e_k)$. Conversely, a finite $V^{(2)}_{res}$ gives reason for finite
self-energies for which both real and imaginary parts are considered
according to the formulae $\Sigma(\omega,T)= 
\Sigma^\prime(\omega,T)-i\Gamma(\omega,T) /2$, with
\begin{equation}
\label{width}
\Gamma(\omega,T)=
    {1\over \Gamma_0}\;{(\hbar \omega-\mu)^2 + \pi^2 T^2 \over 
    1 +{1\over c^2}\left[(\hbar \omega-\mu)^2 + \pi^2 T^2 \right]}
\end{equation}
and $\mu$ being the chemical potential. Then the
$\varrho_k(\omega) $ becomes
\begin{equation}
\label{strength}
\varrho_k(\omega) = {\Gamma(\omega,T) \over
    \left(\hbar \omega -e_k -\Sigma^\prime(\omega,T)\right)^2 + 
    \left({\Gamma(\omega,T)\over 2} \right)^2}
\end{equation}
In (\ref{width}) the $1/\Gamma_0$ represents the strength of the
"collisions", viz the coupling to more complicated states. The
cut-off parameter $c$ allows one to account for the fact that the
imaginary part of the self-energy  does not increase indefinitely when
the excitations get away from the Fermi surface.  In the present
calculation we choose $\Gamma^{-1}_{0}=0.03 MeV^{-1}$ and $c=20 MeV$.
We would like to draw the readers attention to the frequency dependence
put into (\ref{width}). Speaking of the collision term in BUU type
equations once more, the latter would have to account for subtle memory
effects if it should cope with the form we use for the self energies.

So far it was not specified whether or not the sum over $j,k$ should
include diagonal matrix elements. Quite generally, they measure
quasi-static properties of the system and they are responsible for the
"heat pole". The latter shows up at $\omega=0$ either in the relaxation
function or in the correlation function associated to the dissipative
part of the response. It was argued in \cite{hoivya} to neglect
contributions from this heat pole, last not least to force ergodicity
to be given. Within our model this implies to restrict the summation in
eq.(\ref{response}) to non-diagonal matrix elements. Such a restriction
has thus been done also in the present computations. A more elaborate
discussion will be published in \cite{ivhopaya}. 

\subsection{Friction in zero-frequency limit}

Because of its great importance we like to address specifically the
calculation of the friction coefficient, in particular its temperature
dependence. The origin of the latter can be made transparent for the
zero-frequency limit. As we shall see below, numerical evidence tells
one that this limit represents the actual value quite well for not too
small temperatures (c.f. also \cite{yahosa}, \cite{hosaya} and
\cite{ivhopaya}). So let us evaluate $\gamma(0)$ by inserting
eq.(\ref{response}) into eq.(\ref{gamma0}). One 
obtains:
\begin{equation}
\label{gm0}
\gamma(0) = -
\int {d\hbar\Omega \over 4\pi}\;  
{\partial n(\Omega) \over \partial \Omega}
    \sum _{jk } \big\arrowvert F_{jk }\big\arrowvert ^2  
    \varrho_{k}(\Omega)\varrho_{j}(\Omega).  
\end{equation}
This expression can be simplified further using the Sommerfeld
expansion for $n(\Omega)$. To leading order one gets \cite{ho}
\begin{equation}
\gamma(0)\approx
\frac{\hbar}{4\pi}\sum_{jk}|F_{jk}|^{2}\varrho_{j}(\mu)
\varrho_{k}(\mu)
\label{gm0approx}
\end{equation}
Here, the spectral density $\varrho_{k}(\mu)$ is to be calculated at
the frequency $\hbar \omega = \mu$.  According to (\ref{width}) this
implies to use for $\Gamma(\mu,T)$ a constant value, independent of
$\omega$. For such a situation it does not make much sense to use
refinements like the cut-off in the frequency dependence.  Moreover, as
the real part of the self-energy vanishes at $\hbar \omega =
\mu$ (see \cite{ho}) we may use
\begin{equation}
\Gamma(\mu,T)=\frac{1}{\Gamma_{0}}
\pi^{2}T^{2}\approx 0.3 T^{2} 
\label{width1}
\end{equation}
and 
\begin{equation}
\varrho_{k}(\mu)=\frac{\Gamma(\mu,T)}{(e_{k}-\mu)^2+\frac{1}{4}
\Gamma^{2}(\mu,T)}
\label{stapprox} 
\end{equation}
It may be noted that this approximation is very much related to the
typical relaxation time approximation (see \cite{ho}) for the collision
term of the BUU or Landau-Vlasov equation. 

Finally we may note that for the simple model outlined here the
dependence of $\gamma(0)$ on $T$ appears to be synonymous to that on a
constant damping width $\Gamma$. The validity of the Sommerfeld
expansion has been studied numerically by computing eq.(\ref{gm0}) for
several temperatures, and for such a model of a constant $\Gamma$
instead of $\Gamma(\omega,T)$. It was found that the temperature
dependence due to $n(\Omega)$ is weak, which in a sense justifies the
use of (\ref{gm0approx}). Therefore, a simple estimate of the
temperature-dependence of friction may be obtained by using a constant
width $\Gamma$ and by relating the latter to $T$ by way of
(\ref{width1}). The dependence of $\gamma(0)$ on $\Gamma$ may be
studied explicitly within a simple model. Take a schematic nucleus
consisting of A/2 protons and A/2 neutrons, which occupy oscillator
shells with principal quantum numbers $1,2,\ldots,N$ and neglect the
spin-orbit coupling. For quadrupole vibrations around a sphere one may
proceed along the lines described in sect.8.5 of \cite{sije}. Applying this
model to (\ref{gm0approx}) with (\ref{stapprox}) one gets
\begin{equation}
\gamma(0)/\hbar\approx\frac{4\Gamma^{2}(2\hbar\omega_{0})^{2}\sigma}
{9\pi[(\hbar\omega_{0})^{2}+\Gamma^{2}]}[\frac{1}
{(3\hbar\omega_{0})^{2}+\Gamma^{2}}+\frac{1}{(5\hbar\omega_{0})^{2}+\Gamma^{2}}
+\cdots]
\label{gm0nols}
\end{equation}
The quantity $\sigma$ represents the dependence of the average squared
matrix elements on $N$. To leading order one gets
$\sigma\simeq(\frac{3}{2}A)^{\frac{4}{3}}$ (c.f.\cite{sije}).

\subsection{The single-particle Hamiltonian}

For the present computation the single particle Hamiltonian
$\hat{h}(\hat{x},\hat{p},Q)$ of (\ref{hmf}) is chosen to be given by
the two-center shell model:
\begin{equation}
\hat{h}=-\frac{\hbar^{2}\nabla^{2}}{2m}+\hat{V}(\hat{\rho},\hat{z})
+\hat{V}_{ls}(\hat{x},\hat{l},\hat{s})+\hat{V}_{l^{2}}(\hat{x},\hat{l})
\label{twoham}
\end{equation}
This model has been developed some time ago \cite{magr},
(see also \cite{yaiwa}) and has been applied successfully to heavy ion
collisions and fission. It is comparatively simple numerically but
nevertheless is able to describe quite well \cite{schgrmo} the single
particle energies and wave functions near the Fermi level.

\begin{figure}[p]
\centerline{{\epsfysize=10cm \epsffile{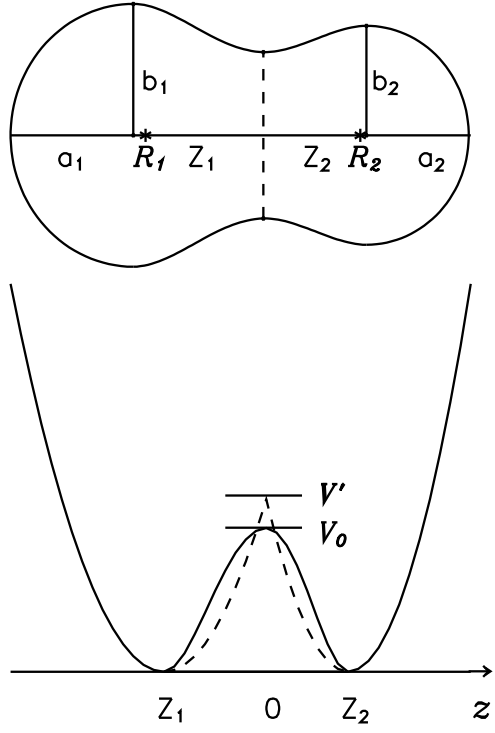}}}
\caption{Lower part: The single particle potential  $\hat{V}(\rho,z)$
along the z-axis; $V_{0}$ and $V'$ are the heights of the potential
barrier at the origin, calculated with and without the neck-correction term,
respectively; Upper part: The associated nuclear shape as specified 
by an equipotential surface.}
\label{shape}
\end{figure}

Here $V$ is the two-center potential in cylindrical coordinates with $m$,
$\hat{s}$ and $\hat{l}$ being the nucleons' mass and the operators for
spin and angular momentum, respectively. A detailed description of 
the construction of $V$ can be found in \cite{magr} and \cite{yaiwa}.
As basic shape parameters there are the distance $z_{0}=z_{2}-z_{1}$
between the centers $z_i$ of the two potential wells,  a neck-parameter
$\epsilon$, a mass ratio $\alpha$ and deformations $\delta_{1}$,
$\delta_{2}$ of the two fragments.  To demonstrate the geometrical
meaning of the shape variables, we show in Fig.\ref{shape} the
potential along the z-axis and the associated nuclear shape. As usual
the nuclear surface is identified as an equipotential one of the
two-center potential, and volume conservation is required for the
uniform density inside the surface corresponding to the Fermi energy.
The volume then has a value given by the sphere with radius
$R_{0}=r_{0}A^{1/3}$, with A being a mass number of the nucleus.  In
the present calculation, we take the radius parameter $r_{0}$ to be
equal to 1.2 fm. The associated oscillator frequency $\hbar \omega_{0}$
for the spherical shape is then given by $\hbar\omega_{0}=41A^{-1/3}$.

The momentum-dependent part in eq.(\ref{twoham}) consist of a spin
orbit-coupling term $\hat{V}_{ls}$ and an $l^{2}$-term
$\hat{V}_{l^{2}}$.  For them the angular momentum $\hat{l}$ is
described in the stretched coordinates.  The strengths $\kappa_{i}$ of
$ls-$ and $\kappa_{i}\mu_{i}$ of $l^2-$ are taken from \cite{schgrmo}.
For a given shape the Hamiltonian (\ref{twoham}) is diagonalized within
the basis of eigen-functions constructed for the the two-center
potential without a "neck-correction term"; hereto states up to
energies of $(N_{0}+3/2)\hbar\omega_{0}$ are taken into account, with
$N_{0}=20$. The "neck-correction term" is used to remove the cusp which
naturally arises at $z=0$ if one just puts together two oscillator
potentials (see Fig.\ref{shape}; for more details we refer to \cite{magr}
\cite{yaiwa}).  The deformations $\delta_{i}$ of the fragments are
described by the ratio of the semi-axes $a_i$ and $b_i$ as
$a_{i}/b_{i}=\sqrt{(1+2/3\delta_{i})}/\sqrt{(1-4/3\delta_{i})}$, with
$i=$ 1 and 2 for the left and right fragments.  The neck-parameter
$\epsilon$ is defined as the ratio $\epsilon=V_{0}/V'$ with $V_{0}$ and
$V'$ being the heights  of the potential barrier at the origin,
calculated with and without the "neck-correction term", respectively.

As mentioned earlier the transport coefficients shall be computed
along a "fission path" only. In this paper we want to identify the
latter with a line of minimal potential energy. This energy shall be
approximated by the one of the cold liquid drop, calculated for a two
dimensional subspace, which comes about in the following way: We
concentrate on symmetric fission, for which the mass ratio $\alpha$
becomes unity, and choose the deformations of the left and right
fragments to be equal to each other, which means to have
$\delta_{1}=\delta_{2}=\delta$.  Finally, the neck parameter, which is
sensitive to specifications of the scission configuration, is fixed to
the value $\epsilon=0.4$. In a previous study \cite{waabe} such a
choice of the "fission path" has been seen to reproduce well the
observed average kinetic energy for thermal fission of $^{200}Pb$.  The
family of shapes from which the "fission path" is evaluated is defined
by the two parameters $z_{0}$ and $\delta$, and the motion along it
shall be parameterized by a variable $r_{12}$. It is defined as the
ratio $r_{12}=R_{12}/(2R_{0})$, where $R_{12}=R_{2}-R_{1}$ measures the
relative distance between the centers of mass of the two fragments (see
Fig.\ref{shape}), and $R_{0}$ stands for the radius of the spherical
configuration. For this variable $r_{12}$ the single-particle operator
$\hat{F}$ is defined as
\begin{equation}
\hat{F}={{\partial \hat h}\over{\partial r_{12}}}=
\left({\frac{\partial r_{12}}{\partial z_{0}}+
\frac{\partial \delta}{\partial z_0}
\frac{\partial r_{12}}{\partial\delta}}\right)^{-1}
\left(\frac{\partial\hat{h}}{\partial z_0}+
\frac{\partial \delta}{\partial
z_0}\frac{\partial\hat{h}}{\partial\delta}\right) 
\label{F_Q}
\end{equation}
where the derivative ${\partial \delta}/{\partial z_0}$ is to be taken along 
the fission path $\delta =\delta (z_0)$.

\subsection{The deformation energy}
For the collective response function eq.(\ref{collresp}), we
need the coupling constant $k$ as given by eq.(\ref{coupling3}).
This expression can be evaluated from the free energy $f$ and the static
response $\chi(0)$. The free energy $f(r_{12},T)$ will be written as a sum of
Coulomb and surface energies plus the shell correction part: 
\begin{equation}
f(r_{12},T)=f_{Coul}(r_{12},T)+f_{surf}(r_{12},T)+f_{sc}(r_{12},T). 
\label{free}
\end{equation}
The free Coulomb and surface energies is approximated by
\begin{equation}
f_{Coul}(r_{12},T)=f_{Coul}(r_{12},T=0)(1-\alpha T^{2})
\label{coul}
\end{equation}
 and 
\begin{equation}
f_{surf}(r_{12},T)=f_{surf}(r_{12},T=0)(1-\beta T^{2})
\label{surf}
\end{equation}
 with the values of 
$\alpha$ and $\beta$ taken to be 0.000763 and 0.00553 $MeV^{-2}$ 
\cite{gustbr}. The free energies $f_{Coul}(r_{12},T=0)$ and
$f_{surf}(r_{12},T=0)$ have been evaluated according to \cite{krnisi}.
In Table III of \cite{krnisi}, there are several sets of parameters.
Here, we choose the set with the values of $a=0.65 fm$, 
$a_{s}=21.836 MeV$, 
$K_{s}=3.48$ corresponding to the radius parameter $r_{0}=1.2 fm$.
For the shell correction we assume the form (c.f.\cite{bomo}):
\begin{equation}
f_{sc}(r_{12},T)=f_{sc}(r_{12},T=0)\tau / \sinh\tau
\label{}
\end{equation}
with
$\tau=(2\pi^{2}T)/(\hbar \omega_{0})$. The shell correction
$f_{sc}(r_{12},T=0)$ is evaluated according to \cite{bofinino}.

\section{Numerical results}

We want to evaluate transport coefficients along the
fission path for symmetric fission of $^{224}Th$ and like to study
their dependence on temperature $T$ and their variation with the shape
variable. As previously mentioned, the fission path will be defined
through the static energy. This is justified in case that the collective
motion is strongly damped, which implies that the system "creeps" down
the potential landscape from saddle to scission. Indeed, later on we
will find this hypothesis justified by the numerical values of the
transport coefficients.

\subsection{The static energy and the stiffness coefficient}

In Fig.\ref{vldm} the liquid-drop energy 
$V_{LDM}(z_{0},\delta)=f_{Coul}(T=0)+f_{surf}(T=0)$ is shown as function of
$z_{0}$ and $\delta$.  The curve along which this
energy is minimal for each $z_{0}$ is given by the
long-dashed curve. It connects the point A of the 
spherical shape with the saddle point C and then follows down to the
scission configuration. The latter part reflects the path of steepest
decent. This trajectory is taken as our "fission path". It is clear
that such a static path will change with temperature,  
because of the dependence on $T$ of both the shell correction as well
as of the  macroscopic free energy. Such refinements are neglected
here, for the following reasons. First of all, we concentrate on
excitations where shell effects are largely gone. Secondly, we know
from numerical experience that the path of minimal potential energy, as
calculated from the "macroscopic part", is not very sensitive to
changes in $T$. 

In Fig.\ref{freeen}, we plot the free energy $f(r_{12},T)$ as a
function of the shape variable $r_{12}$ for different values of $T$
along the fission path shown in  Fig.\ref{vldm}.  We restrict ourselves
to $T\geq 1 MeV$ and neglect pairing correlations, which should be
taken into account for quantitative studies at lower temperatures.  The
second derivative of the free energy $C(0)={\partial^{2}f}/{\partial
Q^{2}_{0}}$ is plotted in Fig.\ref{stiffnesst} as function of $T$.
At not too small temperatures the actual  stiffness $C$ turns out to
be close to this zero-frequency limit; (\ref{capprox}) and the
discussion given there.

\subsection{The friction coefficient}

\leftline{\bf a) A schematic study of the zero frequency limit}
\medskip

We first would like to work out a few general features at the example
of the zero-frequency limit (\ref{gm0approx}) applied to quadrupole
vibrations. In Fig.\ref{zerofric} the result of a computation is shown
for the doubly closed shell nuclei $^{140}Yb$ and $^{208}Pb$,
neglecting spin-orbit coupling in the first case. For the width the
formula $\Gamma(\mu,T)=0.3T^{2}$ is used and as single particle model
a harmonic oscillator is taken with spheroidal deformation $\delta$. 
For small deformations this $\delta$ is related to $\alpha_{2}$ by
$\delta\sim ({15}/{8\pi})\alpha_{2}$, where $\alpha_{2}$ is the common
parameter appearing in  ($R=R_{0}(1+\alpha_{2}Y_{20}(\theta,\varphi))$)
\cite{bomo}, \cite{sije}.  

Fig.\ref{zerofric} shows that at small $T$ friction increases strongly with
$T$, then it reaches some maximal value at some $T_{max}$ and 
decreases afterwards. Taken simply as function of the width $\Gamma$, 
such behavior may be inferred from the semi-analytical formula
(\ref{gm0nols}). With respect to the dependence on temperature one must
be a little careful, as the result shown in the figure was obtained for
$\Gamma(\mu,T)=0.3T^{2}$, neglecting the influence of the cut-off
parameter $c$ in (\ref{width}). A finite cut-off parameter would weaken
the $T$-dependence somewhat (see \cite{hoivya}), but it should be clear
that the present model study only aims at a qualitative understanding.
After all we are neglecting subtle, but important details of
collisional damping like the width's dependence on frequency.  The
value of the width $\Gamma_{max}$ at which $\gamma(0)$ reaches its
maximal value is strongly related to the energy differences
$|e_{j}-e_{k}|$ of  those single-particle pairs $(j,k)$ which give
significant values to the transition matrix element $|F_{jk}|^{2}$.
From eq.(\ref{gm0nols}) one may deduce this $\Gamma_{max}$ to be of the
order of $\hbar\omega_{0}$ in case of no spin-orbit coupling. With
spin-orbit coupling the values of $\Gamma_{max}$ will get smaller,
since then there will be important contributions from  pairs $(j,k)$
which lie closer in energy. The corresponding $T_{max}$ may be
estimated to $0.3\times T_{max}^{2}\approx \Gamma_{max}$.  For example,
it can be seen that the values of $T_{max}$ are 6.7 Mev for $^{140}Yb$
and 2.5 MeV for $^{208}Pb$.

Let us address now the value of the friction coefficient and compare it
with the wall formula. The maximal values of $\gamma(0)/\hbar$ are seen
to be 40 and 180 for $^{140}Yb$ and $^{208}Pb$, respectively. These
numbers reflect a general behavior 
in that friction can be said to become larger if the spin-orbit
coupling is taken into account, for given mass number.  However,
altogether the magnitude is much smaller than that
of the wall formula, for which one would get 530 and 900 for the two
cases considered, namely $^{140}Yb$ and
$^{208}Pb$, respectively. The latter numbers can be estimated as follows.
For the spheroidal deformation $\delta$ the wall friction \cite{bbnrrss} 
\begin{equation}
\gamma_{w}=\frac{3}{4}\rho v_{F}\oint v_{n}^{2}dS
\label{gammaw}
\end{equation}
can be shown to reduce to the simple expression
$\gamma_{w}/\hbar=\frac{2\pi}{25}\sqrt[3]{9\pi}A^{\frac{4}{3}}\approx 
0.76A^{\frac{4}{3}}$ ($\rho$ is the particles density, $v_{F}$ - Fermi
velocity, $v_{n}$ - wall velocity normal to surface).

The small value obtained for our friction coefficient is related 
to the "gap" in the single-particle spectrum, which in the
expressions for the friction coefficient appears in the denominators
(mind (\ref{gm0nols})).  For doubly closed nuclei the contribution to
friction comes from states whose energy difference  $|e_j-e_k|$ is
rather large. This energy difference becomes especially large for the
spherical oscillator potential (for which only states with
$|e_j-e_k|=2\hbar\omega_0$ contributes to response function and
consequently to the friction).

\bigskip
\leftline{\bf b) General results}
\medskip

Let us turn now to discuss results which are obtained using our
complete theoretical input, namely the realistic two center
shell model, applied to expression (\ref{response}), with the collisional
damping as given by (\ref{strength}) and (\ref{width}), together with
the realistic coupling constant. The coefficient $\gamma$ is shown in
Fig.\ref{gammat} as function of temperature, for the points A and C
along the fission path indicated in Fig.\ref{vldm}. For comparison
on the right the values of the so called modified wall formula
are shown as well, which reads
$\gamma_{m.w}=0.27\gamma_{w}$\cite{nisi}. 

Two computations are presented, one based on the fit of the oscillator
response (see (\ref{oscresp}) and one based on the formula
(\ref{gmapprox}) (with $C(0), \chi(0)$ and $\gamma(0)$ being calculated
from the original $\chi(\omega)$, without employing any fit). It is
seen that both results agree almost perfectly well for all temperatures
used at those points where the local stiffness is negative, and where,
hence, the local motion is unstable.  At the stable point A there is
some disagreement at smaller temperatures, but even at $T=1 \; MeV$ the
difference is smaller than about $30 \%$ and at  $T=1.5 \; MeV$ it is
almost negligible.  The origin of the deviation of these two
computations from each other can easily be understood looking at the
collective strength distributions discussed above with the help of
Fig.\ref{imagcoll}. It is only for small temperatures that this
distribution cannot be approximated by one Lorentzian. We should like
to note that the friction coefficient given by (\ref{gmapprox}) is
almost identical to the one of the zero-frequency limit. The correction
given by $({C(0)}/{\chi(0)})^{2}$ only amounts to about $4\%$.

Finally, we exhibit in Fig.\ref{gammaq} the coordinate dependence of
the friction coefficient for various temperatures. It so turns out that for
the present case the result at $T=1\; MeV$ is very well represented by
the modified wall formula.

\subsection{Mass parameter}

Finally we turn to the inertia for which some new problems appear, as
may be demonstrated with the help of Fig.\ref{compmass}.
There the temperature dependence is exhibited for three computations of
the inertia, performed at the point A. Firstly, there is the one from
the fit and secondly the one calculated from the expression
(\ref{msapprox}). Like for the case of friction, both agree with each
other for higher temperatures, say above $1.5 \; MeV$, and for the same
reason. Above this temperature essentially no higher modes exist,
at least not for the present model. It is important to note that,
different to the case of friction, for the inertia the zero-frequency
limit $M(0)$, as given by the "cranking inertia" (\ref{mass0}) fails to
be a reasonable approximation.  Looking at (\ref{msapprox}) it is only
by way of the second term in the second bracket that the $M$ calculated
this way is close to the one of the fit. As has been noted before, the
zero-frequency mass may become very small, often even negative.

Let us address now the third curve of Fig.\ref{compmass}, which is
marked by "sum rule". As is well known from the case of zero damping,
see e.g. \cite{bomo} or \cite{sije}, there should exist the following,
general relation of the energy weighted sum to the inertias of the
various possible modes with frequencies $\omega_\nu^\pm$ (see also
\cite{ho}): 
\begin{eqnarray}
S=\langle[\hat{F},[\hat{H},\hat{F}]]\rangle
=\frac{\hbar^{2}}{\pi}\int^{\infty}_{-\infty}d\omega
\chi^{\prime\prime}_{coll}(\omega)\omega  \nonumber \\
=\sum_{\nu} \frac{\hbar^{2}}{\pi}
\int^{\infty}_{-\infty}d\omega\chi^{\prime\prime (\nu)}_{osc}(\omega)\omega
=\sum_{\nu}\frac{\hbar^{2}}{k^{2}M(\omega_{\nu})}
\label{sumrule}
\end{eqnarray}
(Here, in the last expression again the factor $k^2$ appears because we
take the inertia of the $Q$-mode but relate it to the sum of the
$F$-modes). As may be inferred easily from this formula, the inertia of
one mode cannot be smaller than $\hbar^2 /( S k^2)$. Unfortunately,
this is not necessarily the case without further precautions.

Essentially two problems appear here. First of all, evaluating
the integral
\begin{equation} 
S=2\frac{\hbar^{2}}{\pi}\int^{\omega_{max}}_{0}d\omega
\chi^{\prime\prime}_{coll}(\omega)\omega
\label{sum}
\end{equation}
as function of the upper limit $\omega_{max}$ one realizes bad
convergence. This fact is related to the construction of the response
function, particularly that of $\chi^{\prime\prime}$. The forms given
in (\ref{strength}) and (\ref{width})  cannot be expected to be good at
large frequencies.  The second problem comes up whenever the single
particle potential does not warrant the density of the shell model to
agree sufficiently well with the one which is used in the definition of
the shape variable. Usually, this is all right for the simple model of
the deformed oscillator, for which it is easy to prove that the inertia
corresponding to the total sum, i.e. $\hbar^2 /( S k^2)$, is given by
the $M_{irrot}$ irrotational flow. Indeed, for this case it has been
possible in \cite{hoyaje} to show that the self-consistent inertia
$M(\omega_1)$ turns into $M_{irrot}$ for large temperatures. Again, the
very fact that $M(\omega_1)$ must approach the value given by the sum
rule follows from the observation that with increasing excitation the
strength concentrates in the low frequency mode.

In a forthcoming paper \cite{ivhopaya} we will try to solve the second
problem mentioned above, which involves the relation of the  
density distribution to the shape of the potential. For the present
case we suggest the following pragmatic procedure. We just fix the sum
rule value by the inertia $M_{irrot}$ of irrotational flow, which in
turn is evaluated by the method of Werner-Wheeler \cite{dasini}. To put
it differently, we scale the inertia $M=M(\omega_1)$ for the low
frequency mode such that it becomes equal to $M_{irrot}$ at large $T$.
At smaller temperatures this $M$ will attain larger values, of course.
The result of such a computation is shown by that curve which in
Fig.\ref{compmass} is marked by "sum rule".

In this context it is worth recalling that at somewhat larger
temperatures collective motion gets over-damped such that inertia
does not play a big role anymore. It is reassuring therefore to see
that for such cases the fit of the oscillator response does not depend
very much on the inertia (mind the discussion given above next to 
(\ref{ommax})). 

In Fig.\ref{massq} we address the coordinate dependence, as obtained
for various temperatures. From this figure one may observe that above
$T= 2 \;MeV$ the values do not change anymore with excitation.

\subsection{The inverse relaxation time $\beta=\gamma/M$}

In the literature the quantity  $\beta=\gamma/M$ has come into use 
(see e.g.\cite{hirorev} to \cite{pauthoe}) for
parameterizing the strength of the friction force, where for $M$
commonly the irrotational flow value is taken. The inverse quantity
$\beta^{-1}$ has a dimension of time and actually measures physically
the (local) relaxation time to the Maxwell distribution in collective
phase space.  Taking our results for friction and inertia from
Figs.\ref{gammaq} and \ref{massq}  we get a coordinate dependence of
$\beta$ as shown in Fig.\ref{betaq}. It is seen that this dependence is
much weaker than the one of $\gamma$ and $M$ themselves. This is to be
expected in a sense, as any geometrical factor contained in the
individual quantities drops out when building the ratio. Actually, the
variation along the fission path is weaker than the one with
excitation. Incidentally, it is seen that $\beta$ decreases somewhat
with deformation.

The values of $\beta$ seen in Fig.\ref{betaq} are in the range of those
published earlier for quadrupole vibrations of $^{208} Pb$
\cite{hoyaje} \cite{hoivya}, having the tendency of being slightly
smaller than in this previous case, which is no surprise in the light
of the discussion presented above for friction.  In any case, these
values are comparable to those associated with "linear response theory"
in the compilation given in Fig.13 of \cite{higoroobn}. As seen from
Fig.12 of the same paper, they are in reassuring agreement with numbers
found adequate in many theoretical descriptions of fission accompanied
by emission of light particles.

In Fig.\ref{betaq} our results for $\beta$ are compared with that
suggested by Fr\"{o}brich and Gontchar for a phenomenological description of
fission of the type just mentioned \cite{froe}. Their deformation
parameter $q$ is the same as our parameter $r_{12}$. It may be said
that their model has also been quite successful in explaining observed
experimental data, but quite apparently, their $\beta(q)$ is in violent
contradiction to our picture. In case it will turn out that their model
is safe otherwise, leaving no room for modifying the $\beta(q)$, one
might be inclined to conclude that the friction necessary in this case
must be of different physical origin. One possible candidate would be a
dissipation mechanism which comes from the so called "heat pole" and
which in \cite{hoivya} has been seen to be identical to the one of
"diabatic" motion \cite{aynoediabfri}. This would be an interesting
perspective as it might indicate that motion beyond the saddle is
"fast", even at high excitations, in contradiction to the general
believe.

\subsection{The effective damping rate
$\eta=\gamma/(2\protect\sqrt{M|C|})$ and the dynamics of fission}

For a damped oscillator the quantity $\eta=\gamma/2\sqrt{M |C|}$
indicates whether motion is under-damped ($\eta <1$) or over-damped
($\eta>1$). Therefore, it is a good measure for the effective degree of
damping, except around points of inflection where the local stiffness
becomes zero. Furthermore, it is exactly this quantity, if calculated
at the saddle point, which according to Kramers' famous formula
\begin{equation} 
\label{kramers}
{\rm R_K} = \left(\sqrt{1 + \eta_s} - \eta_s\right) \;
{\varpi_m \over 2\pi}\, e^{-B/T} \approx
\left(\sqrt{1 + \eta_s} - \eta_s\right) \;{\rm R_{BW}}
\end{equation}
determines the deviation of the fission decay rate ${\rm R_K}$ from the
${\rm R_{BW}}$ of the Bohr-Wheeler formula ($\varpi^2=\vert C \vert
/M$). Within the linear response approach $\eta$ has been calculated as
function of $T$ before (see \cite{hoivya}), but only for the case of
quadrupole vibrations around the spherical minimum of lead. From all
the information presented above we are now in the position of
calculating $\eta $ along the fission path. However, in the light of
formula (\ref{kramers}) it may suffice first to study it around the
barrier.  Indeed, this formula applies if, from the full dynamics
across the barrier, it is only the motion around the stationary points
which matters. The latter situation can be seen to be given if the
height $B$ of the barrier is much larger than temperature: $B/T \gg 1$
(\cite{kram}, see also \cite{hofthoming}). Effectively, one may then
approximate the true barrier by a model potential which consists of one
ordinary oscillator for the minimum plus an inverted one for the
barrier, joined smoothly to each other (see \cite{hofthoming}). 
Unfortunately, often (\ref{kramers}) is applied even in cases that
this basic condition on $T$ and $B$ is violated. 

To study these questions for our example we have calculated all
relevant quantities by performing averages both around the potential
minimum as well as around the barrier. For the first case this means to
average in the regime from $r_{12}=0.375$  to some value
$r_{12}^{(1)}$, and for the second one from this $r_{12}^{(1)}$ to some
$r_{12}^{(2)}$ behind the barrier. Both values depend on temperature,
of course, as may be inferred from Fig.\ref{freeen}. We took the values
$r_{12}^{(1)}=0.47,\;0.46,\;0.45,\;0.44,\;0.42$  and
$r_{12}^{(2)}=1.05,\;0.96,\;0.87,\;0.74,\;0.60$  for $T=
1,\;2,\;3,\;4,\;5$ MeV, respectively. The results are put together in
the following Table:

$\begin{array}{cccccccccc}
T & C_m & \hbar \varpi_m & \beta_m & \eta_m &
-C_s &B& \hbar \varpi_s & \beta_s & \eta_s \\
MeV&MeV&MeV&MeV/\hbar&&MeV&MeV&MeV&MeV/\hbar& \\
1 & 538 & 0.82 & 1.8 & 1.1 & 109 & 7.20 & 0.50 & 1.2 & 1.2 \\
2 & 509 & 0.94 & 4.2 & 2.2 & 131 & 5.95 & 0.59 & 3.2 & 2.7 \\
3 & 472 & 0.93 & 4.8 & 2.5 & 134 & 4.31 & 0.59 & 3.8 & 3.2 \\
4 & 407 & 0.87 & 4.6 & 2.7 & 152 & 2.59 & 0.60 & 4.0 & 3.3 \\
5 & 335 & 0.79 & 4.6 & 2.9 & 200 & 1.18 & 0.67 & 4.1 & 3.0 
\end{array}
$
\\Several inferences can be drawn immediately:
\\{\bf -} In the values of $\eta$ and $\beta$, as well as in
their variation with $T$ there is not much difference between the
barrier and the minimum.
\\{\bf -} The basic condition on the validity of
(\ref{kramers}) is satisfied for temperatures up to 3 MeV at best.
\\{\bf -} It is interesting to note that $\hbar\varpi_s$ 
increases with $T$, albeit only slightly.
\\ A closer inspection shows the quantity $2\pi T/(\hbar\varpi_s)$ to
be definitely larger than unity. This implies that collective quantum
effects do not contribute to fission dynamics for the examples chosen
here. According to \cite{hofthoming} for  $2\pi T/(\hbar\varpi_s) \ll
1$ such effects would change (\ref{kramers}) to ${\rm R}={\rm
f_Q}\,{\rm R_K}$, but quite apparently the quantum correction factor
${\rm f_Q}$ plays a role only at temperatures smaller than 1 MeV.

\section{Summary, conclusion and outlook}

In the previous sections we have presented a detailed microscopic
study of transport coefficients for fission. We concentrated on average
motion as in the range of temperatures considered the Einstein relation
holds true. The latter allows one to deduce the essential diffusion
coefficient from friction: $D= \gamma T$. Those for average motion,
$M,\gamma$ and $C$, have been deduced within the quasi-static picture in
which linear response theory is applied to describe motion around a local
thermal equilibrium. Following earlier work, the contribution from the
"heat pole" to the response functions has been neglected. As described
in \cite{hoivya}, such a restriction is closely related to what in the
context of nuclear physics one commonly associates with the "adiabatic"
picture, in contrast to the "diabatic" one one would expect to apply
for "fast" motion.

The response functions have been calculated within a realistic
two-center shell model. Effects of collisions have been accounted for
by using self-energies having both real and imaginary parts. The latter
are allowed to depend not only on temperature but on frequency as
well, in which way memory effects of the collision term are simulated. 
Locally collective motion is treated self-consistently, in the sense
that the structure of the associated response resembles the one known
from the RPA of undamped motion.
 
Various procedures have been described to deduce transport coefficients
from the microscopically computed response, which generalize
previous descriptions. They have been evaluated as function both of
temperature as well as of the variation of the shape along the fission
path. Considering all the effects taken into account, such a study has
never been reported before. The results found invite us to draw the 
following conclusions. 
\\ $\bullet$ Perhaps one of the most striking features is the weak
dependence of both $\beta$ as well as of $\eta$ on the shape parameter, as
encountered here along the fission path. 
\\ $\bullet$ The temperature dependence of these two quantities is
similar to the one reported previously in \cite{hoivya} for vibrations
of  $^{208} Pb$: they increase with $T$, to eventually reach some
saturation around $T\approx 4 \; MeV$.
\\ $\bullet$ This behavior is in agreement with findings reported in
\cite{pauthoe}, at least qualitatively. Our damping rate is somewhat
smaller, and its variation with the shape is weaker than needed in
\cite{pauthoe} for an the analysis of the 
$\gamma$-ray multiplicity encountered for fission of 
$^{224}Th$.
\\ $\bullet$ However, it must be said that we have not yet striven
for quantitative agreements. For instance, there is room for a more
appropriate choice of the two parameters of collisional damping. In the
present paper only their "standard choice" has been used, but as
discussed in \cite{ho} these values are open for changes within a
certain margin, and as demonstrated both in \cite{yahosa} and in
\cite{hoivya}, such modifications will change somewhat (to less than a
factor of 2) the values of the transport coefficients as well as their
variation with $T$. Furthermore, one should mention the influence of
pairing, which still might be important at the smaller $T$ value of 1
MeV, but which we have left out for the sake of simplicity. Moreover,
the influence of angular momentum ought to be taken into account (see
below). It should be said, of course, that the numbers extracted from
comparisons with experiment are model dependent.  Moreover, in
simplified studies often Kramers' picture is applied, even in cases
where the barrier is too low to guaranty that the flux across the
barrier can be described by Kramers' solution of the Fokker-Planck
equation. This problem can only be cured by performing genuine
dynamical studies along the lines reported in
\cite{froe}, \cite{waabe}, \cite{strudiepomsum}, \cite{wada}, \cite{pobaridi}  
but where all ingredients of the macroscopic descriptions with
Fokker-Planck or Langevin equations come from the same theory.
\\ $\bullet$ Finally, we like to come back to angular momentum, once
more, which influences fission dynamics in a manyfold way. First of all,
it changes the quasi-static energy through the centrifugal potential.
Generally, this reduces the height of the barrier and thus
restricts further the range of temperatures for which (\ref{kramers})
is applicable. The centrifugal force will also modify the
local stiffness and thus the coupling constant $k$ through
(\ref{coupling3}), which in turn effects the local response in various
ways. However, even in zero frequency limit rotations will have sizable
impact on the transport coefficients. This has been demonstrated in
\cite{pomhof} for the case of friction. The reason for such a behavior
is found in the fact that in the rotating frame the level structure may
become very different from the one without rotations.

\bigskip
\leftline{\bf Acknowledgments} 
\medskip
The authors want to thank A.S. Jensen for valuable discussions and
suggestions, as well as the Deutsche Forschungsgemeinschaft for
financial support. Two of us (S.Y. and F.A.I.) like to thank the
Physics Department of the TUM for the hospitality extended to them
during their stay.

%


\vskip 1cm
\begin{figure}[p]
\bigskip\bigskip
\caption{The cold liquid-drop energy surface  as functions of 
$z_{0}$ and
$\delta$ for symmetric fission of $^{224}Th$.
The long- and short-dashed curves show the minimal
energy fission path and the configuration where two fragments start
to separate, respectively.
(\protect{\it There exists no postscript file for this figure; the
authors will be glad to send a hard copy upon request})}
\label{vldm}
\end{figure}
\begin{figure}[p]
\centerline{{\rotate[r]{\epsfysize=13cm \epsffile{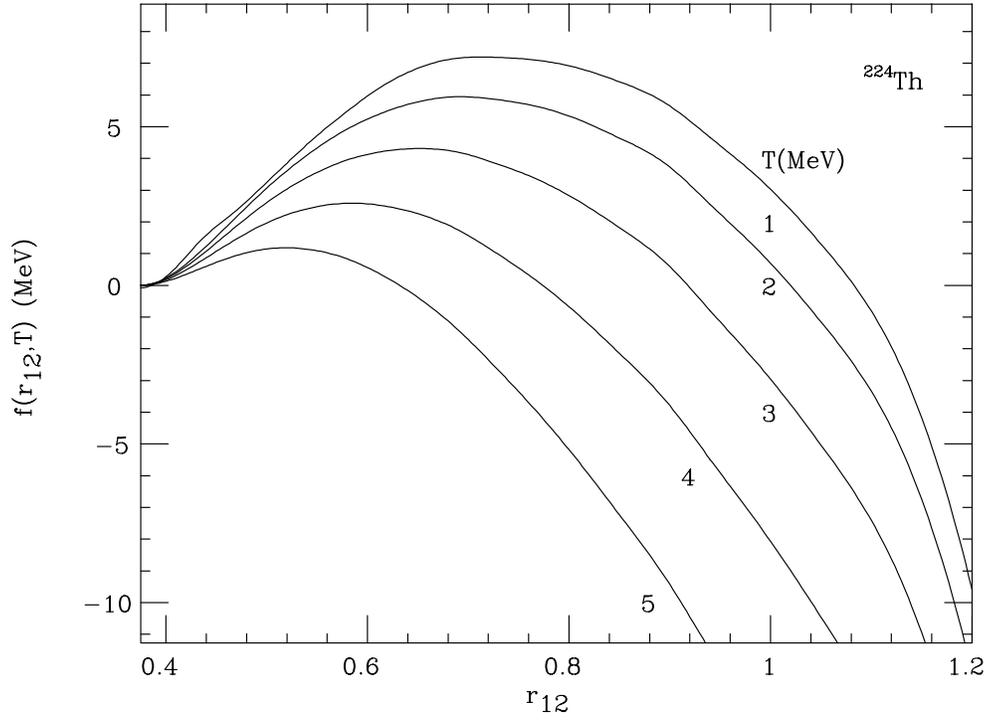}}}}
\caption{The free energy as a function of the shape variable $r_{12}$
along the fission path for temperatures of 1,2,3,4, and 5 $MeV$.} 
\label{freeen}
\end{figure}
\begin{figure}[p]
\centerline{{\rotate[r]{\epsfysize=13cm \epsffile{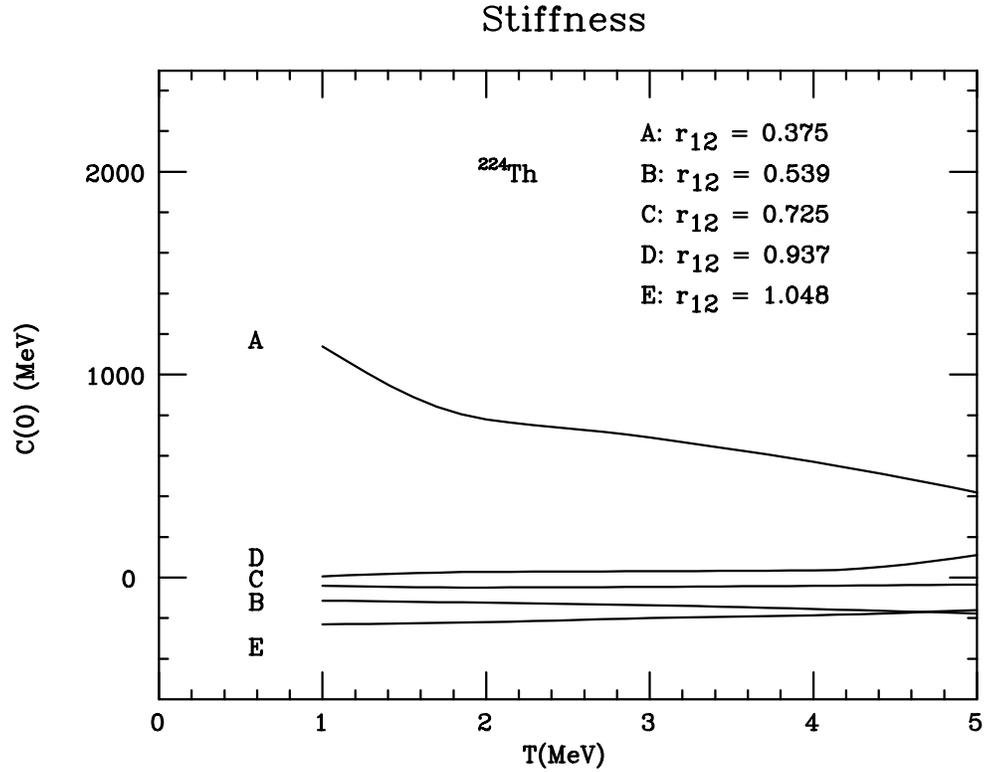}}}}
\caption{The local stiffness $C(0)$ of the zero frequency limit as a
function of $T$ at the points A, C 
}  
\label{stiffnesst}
\end{figure}
\begin{figure}[p]
\centerline{{\rotate[r]{\epsfysize=10cm \epsffile{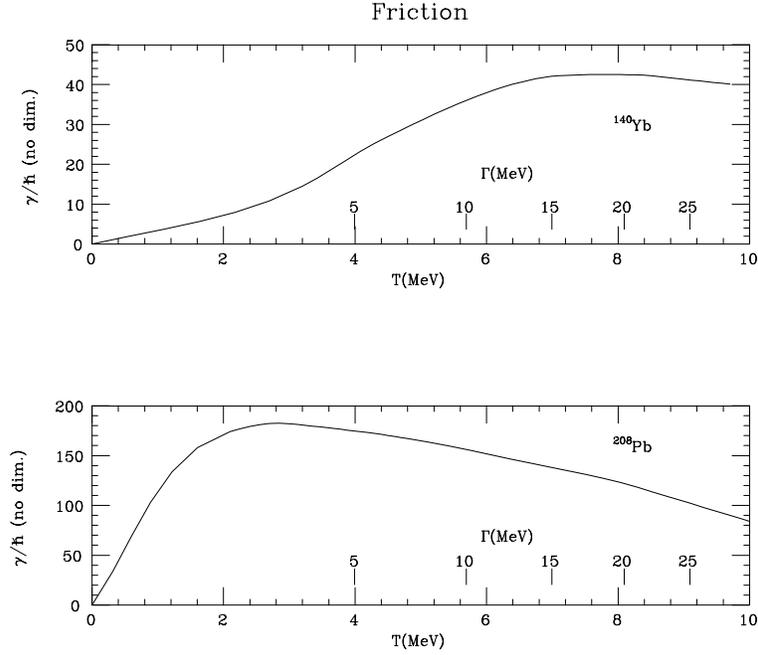}}}}
\caption{The temperature dependence of the friction coefficient 
Eq.(\protect\ref{gm0approx}) 
in the zero-frequency limit  computed with the
oscillator well. The shape variable $Q$ corresponds to $\delta$. The
computations are performed for the double magic nuclei $^{140}Yb$
without spin-orbit coupling (upper) and $^{208}Pb$ 
with spin-orbit coupling (lower)} 
\label{zerofric}
\end{figure}
\begin{figure}[p]
\centerline{{\rotate[r]{\epsfysize=10cm \epsffile{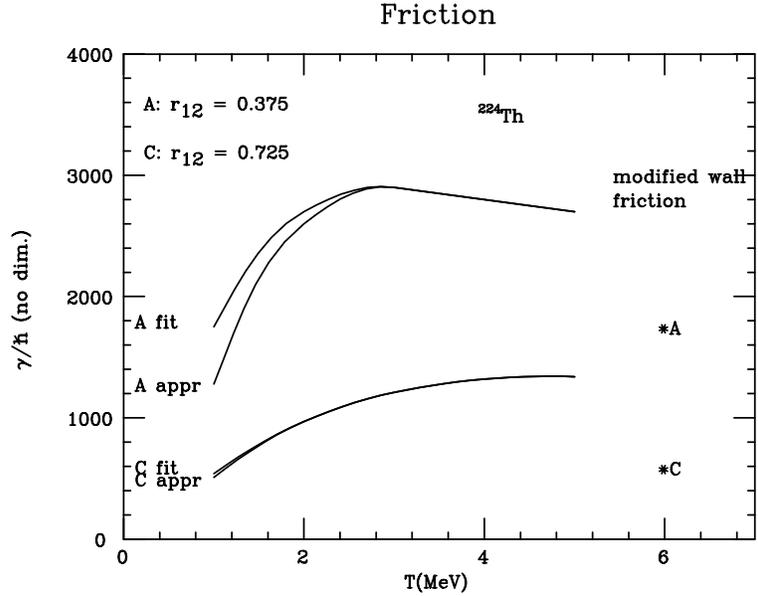}}}}
\caption{The friction coefficient $\gamma$ obtained from the
oscillator fit, shown as function of $T$ at the points A-E
}
\label{gammat}
\end{figure}
\begin{figure}[p]
\centerline{{\rotate[r]{\epsfysize=13cm \epsffile{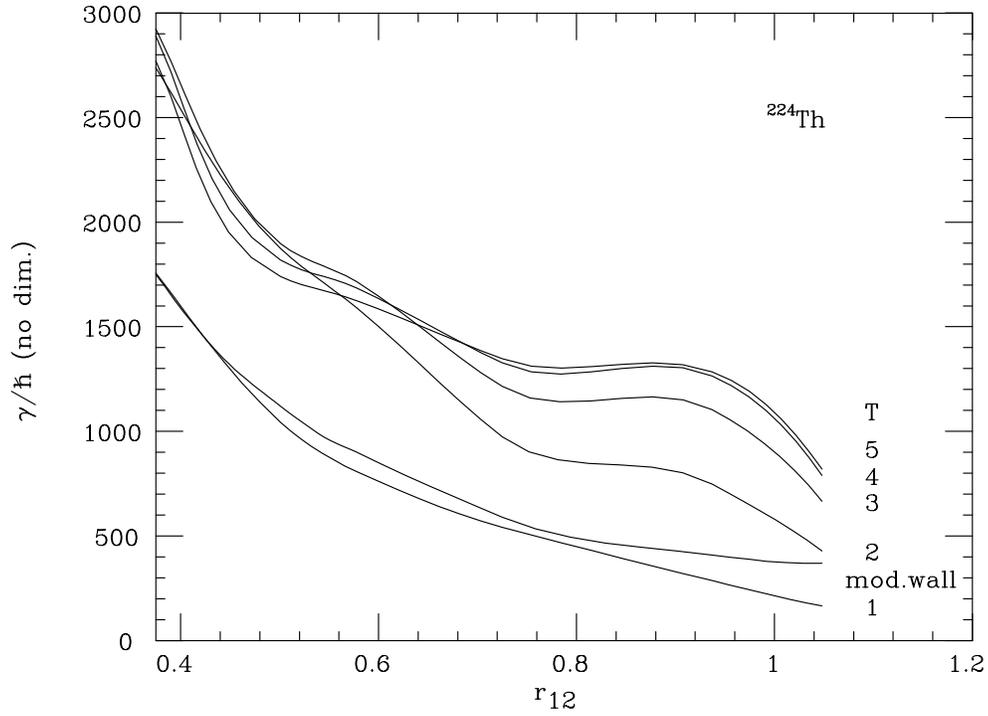}}}}
\caption{The friction coefficient $\gamma$ as a
function of $r_{12}$ for $T=$1-5 $MeV$} 
\label{gammaq}
\end{figure}
\begin{figure}[p]
\centerline{{\rotate[r]{\epsfysize=13cm \epsffile{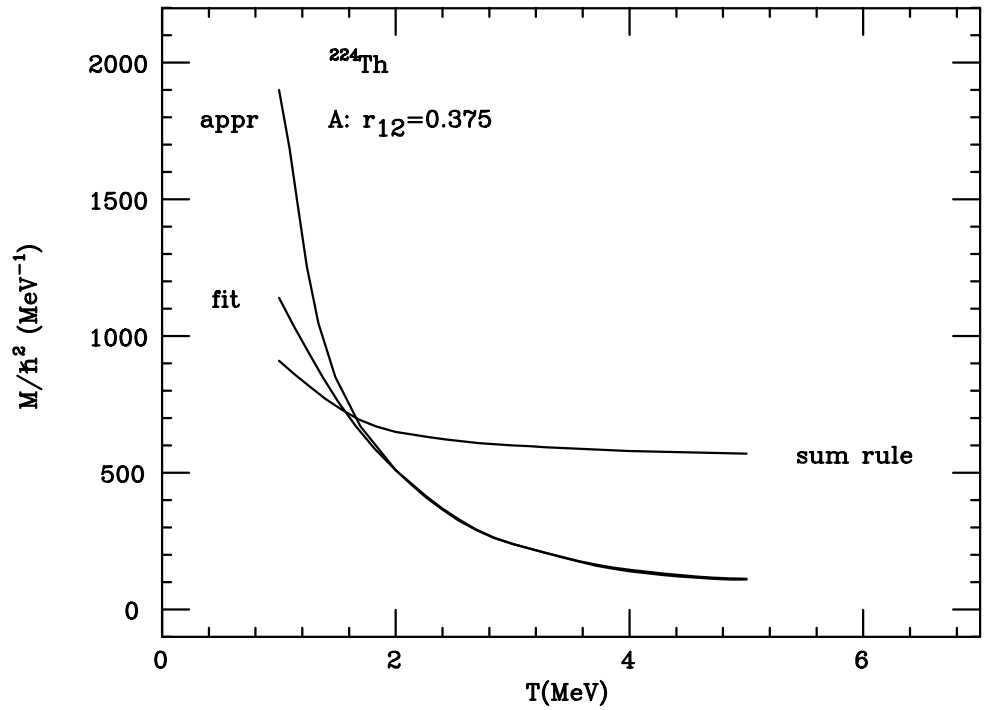}}}}
\caption{The inertia as a function of $T$ at the point A, calculated by three
different methods: (i) by the oscillator fit, (ii) by way of
Eq.(\protect\ref{msapprox}) and (iii) after obeying the sum rule (see text).}
\label{compmass}
\end{figure}
\begin{figure}[p]
\centerline{{\rotate[r]{\epsfysize=12cm \epsffile{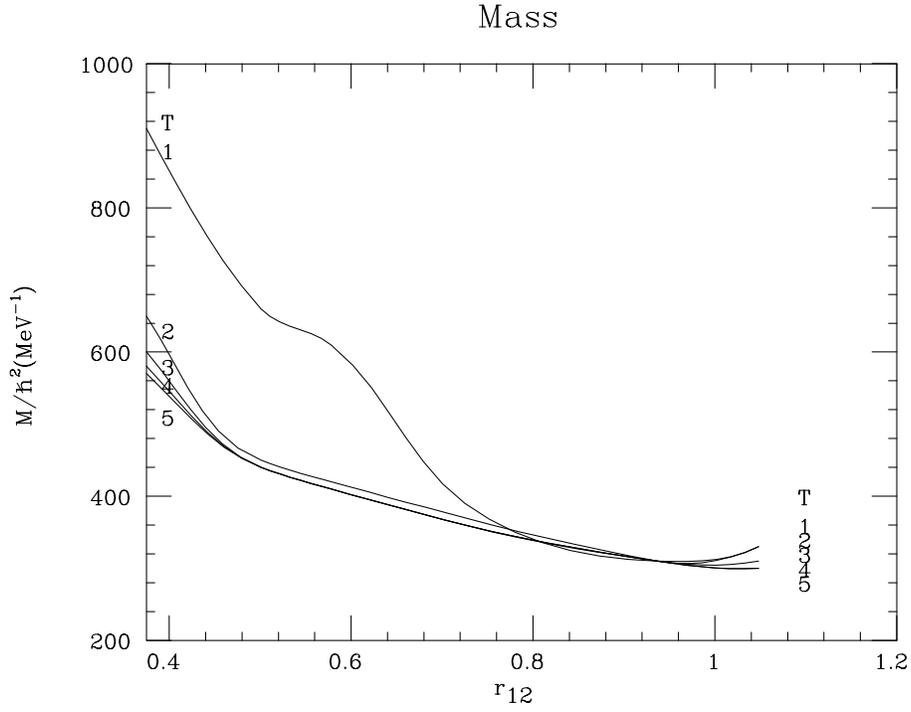}}}}
\caption{The inertia $M$ satisfying the energy weighted sum
rule as a function of $r_{12}$ for $T=$1-5 $MeV$} 
\label{massq}
\end{figure}
\begin{figure}[p]
\centerline{{\rotate[r]{\epsfysize=12cm \epsffile{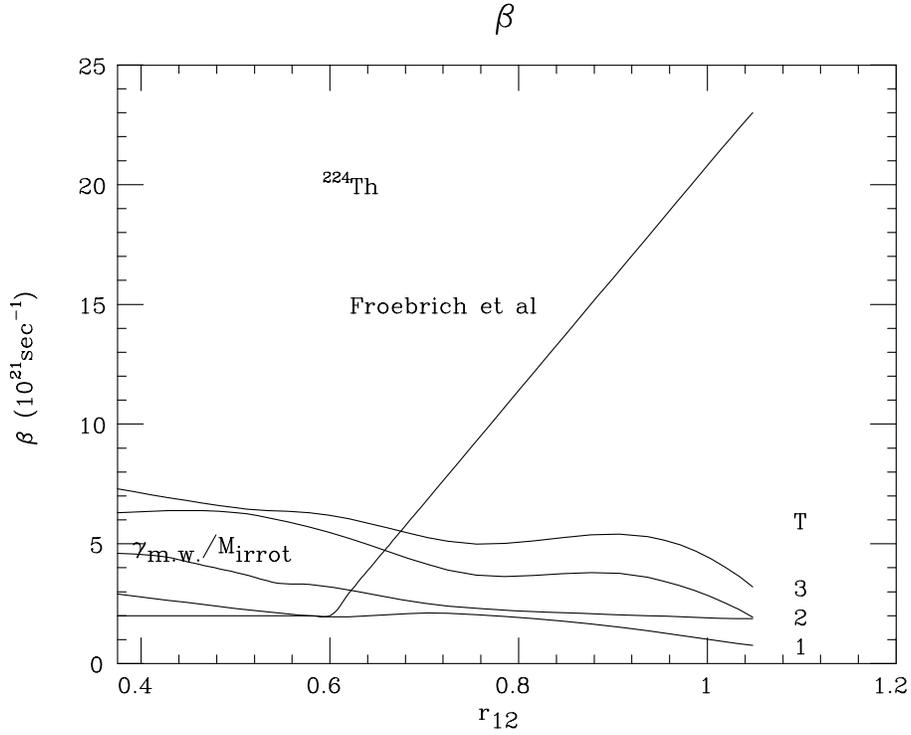}}}}
\caption{$\beta=\gamma/M$ as function of $r_{12}$ for $T=$1-3
$MeV$; for comparison we plot $\beta=\gamma_{m.w}/M_{irrot}$ and the
empirical values from the analysis of Fr\"{o}brich et. al.}
\label{betaq}
\end{figure}

\vfill


\begin{thebibliography}{99}
\bibitem{hirorev} D.~Hilscher and H.~Rossner, 
 Ann. Phys. Fr. {\bf 17} (1992) 471
\bibitem{higoroobn} D. Hilscher, I.I. Gontchar and H. Rossner, 
 Physics of Atomic Nuclei {\bf 57} (1994) 1187-1199
\bibitem{pauthoe} P.~Paul and M.~Thoennessen,
 Ann. Rev. Part. Nucl. Sci. {\bf 44} (1994) 65
\bibitem{bbnrrss}J.Blocki, Y.Boneh, J.R.Nix,  J.Randrup, M.Robel,
A.J.Sierk, W.J. Swiatecki, Ann. Phys. {\bf 113} (1978) 330
\bibitem{swiatcop} W.J.~Swiatecki, Nucl. Phys. {\bf A574} (1994) 233c
\bibitem{yahosa} S.~Yamaji, H.~Hofmann and R.~Samhammer,
 Nucl. Phys. {\bf A475} (1988) 487
\bibitem{hosaya} H.~Hofmann, R.~Samhammer and S.~Yamaji,
 Phys. Lett. {\bf B229} (1989) 309 
\bibitem{hoivya} H.~Hofmann, F.A.~Ivanyuk and S.~Yamaji,
 Nucl. Phys. {\bf A598} (1996) 187
\bibitem{ho} H.~Hofmann, submitted to Physics Reports
\bibitem{kidhofiva} D.~Kiderlen, H.~Hofmann and F.A.~Ivanyuk,
 Nucl.Phys. {\bf A550} (1992) 473
\bibitem{sije}P.J.~Siemens and A.S.~Jensen,"Elements of Nuclei:
 Many-Body Physics with the Strong Interaction", Addison and Wesley, 1987
\bibitem{holet} H. Hofmann, Phys. Lett. {\bf 61B} (1976) 423
\bibitem{bomo} A.~Bohr and B.R.~Mottelson,
 Nuclear Structure, Vol.II (Benjamin,London,1957)
\bibitem{brenigtwo} W.~Brenig, "Statistical Theory of Heat, Nonequlibrium
 Phenomena", Springer, 1989, Berlin
\bibitem{hoyaje} H.~Hofmann, S.~Yamaji and A.S.~Jensen,
 Phys. Lett. {\bf B286} (1992) 1
\bibitem{ivhopaya} F.A.~Ivanyuk, H.~Hofmann, V.V.~Pashkevich and S.~Yamaji,
 to be published
\bibitem{magr} J.~Maruhn and W.~Greiner, Z.f.Phys. {\bf 251} (1972) 431
 161 
\bibitem{schgrmo} D.~Scharnweber, W.~Greiner and U.~Mosel,
 Nucl.Phys. {\bf A164} (1971) 257
\bibitem{yaiwa} S.~Yamaji and A.~Iwamoto, Z.f.Phys. {\bf A313} (1983)
\bibitem{waabe} T.~Wada, Y.~Arimoto, M.~Ohta, A.~Abe and S.~Yamaji,
 2nd RIKEN-INFN Joint Int. Conf. on ``Heavy Ion Collisions''(Wako),
 World Scientific, Singapore, (1995)
\bibitem{gustbr} C.~Guet, E.~Strumberger and M.~Brack,
 Phys.Lett. {\bf B205} (1988) 427
\bibitem{krnisi} H.J.~Krappe, J.R.~Nix and A.~Sierk,
 Phys.Rev. {\bf C20} (1979) 992
\bibitem{bofinino} M.~Bolsterli, E.O.~Fiset, J.R.~Nix and J.L.~Norton,
 Phys. Rev. {\bf C5} (1972) 1050
\bibitem{nisi} J.R.~Nix and A.~Sierk, Preprint LA-UR-86-698 (1986)
\bibitem{dasini} K.T.R.~Davies, A.J.~Sierk and J.R.~Nix,
 Phys.Rev. {\bf C13} (1976) 2385
\bibitem{froe} P.~Fr\"{o}brich and I.I.~Gontchar,
 Nucl.Phys. {\bf A563}, (1993) 326
\bibitem{aynoediabfri} S. Ayik and W. N\"orenberg, 
 Z. Phys. {\bf A309} (1982) 121 
\bibitem{kram} H.A. Kramers, Physica 7 (1940) 284
\bibitem{hofthoming} H. Hofmann, G.-L. Ingold and M. Thoma, Phys. Lett.B
    317 (1993) 489
\bibitem{strudiepomsum} E. Strumberger, K. Dietrich and K. Pomorski,
     Nucl. Phys. A529 (1991) 522
\bibitem{wada} T.~Wada, A.~Abe and N.~Carjan,
  Phys.Rev.Lett. {\bf 70} (1993) 3538
\bibitem{pobaridi} K. Pomorski, J. Bartel, J. Richert
   and K. Dietrich, submitted to Nucl. Phys. A
\bibitem{pomhof} K.~Pomorski and H.~Hofmann,
  Phys.Lett. {\bf B263} (1991) 164
\end{thebibliography}
\end{document}